\documentclass[12pt]{article}
\usepackage{epsfig}

\usepackage{amssymb}
\def\ba{\begin{array}{c}}
\def\ea{\end{array}}

\def\ben{$$}
\def\een{$$}
\newcommand{\bbr}{\br\!\br}

\newcommand{\pbr}{\prec\,}
\newcommand{\pkt}{\,\succ\,\,}
\newcommand{\kt}{\rangle}
\newcommand{\br}{\langle}
\def\bea{\begin{eqnarray}}
\def\eea{\end{eqnarray}}

\def\beax{\begin{eqnarray*}}
\def\eeax{\end{eqnarray*}}

\begin{document}

\titlepage

\vspace{.35cm}

 {\Large \bf

\flushright

        Time-Dependent and/or Nonlocal 
    Representations of Hilbert Spaces
     in Quantum Theory

 }

\vspace{3mm}

 {\large

\flushright

  M. Znojil\footnote{Miloslav Znojil, DrSc., phone: +420 266 173 286, e-mail: 
znojil@ujf.cas.cz, address: Nuclear Physics Institute, Academy of 
Science of the Czech Republic, 250 68 \v{R}e\v{z}, Czech Republic}

 }

\vspace{3mm}

\subsection*{Abstract}

A few recent innovations of applicability of standard textbook 
Quantum Theory are reviewed. The three-Hilbert-space formulation of 
the theory (known from the interacting boson models in nuclear 
physics) is discussed in its slightly broadened four-Hilbert-space 
update. Among applications involving several new scattering and 
bound-state problems the central role is played by the models using 
apparently non-Hermitian (often called ``crypto-Hermitian") 
Hamiltonians with real spectra. The formalism (originally inspired 
by the topical need of mathematically consistent description of 
tobogganic quantum models) is shown to admit even certain unusual 
nonlocal and/or ``moving-frame" representations ${\cal H}^{(S)}$ of 
the standard physical Hilbert space of wave functions.

\subsection*{Keywords}

Quantum Theory, cryptohermitian operators of observables, stable
bound states, unitary scattering, quantum toboggans, supersymmetry,       
time-dependent models

\section{Introduction}

Fourier transformation ${\cal F}: \psi(x) \to \tilde{\psi}(p)$ of 
wave functions converts differential kinetic-energy operator $K\sim 
d^2/dx^2$ into a trivial multiplication by a number, 
$\tilde{K}={\cal F}K{\cal F}^{-1} \sim p^2$. This means that for  
certain quantum systems the Fourier transformation offers a 
simplification of the solution of Schr\"{o}dinger equation.  The 
generalized, nonunitary (often called Dyson) mappings $\Omega$ play 
the same simplifying role in the context of nuclear physics 
\cite{Geyer}. In our present brief review paper we intend to recall 
and discuss very recent progress and, mainly, a few of our own 
results in this direction.

Our text will be more or less self-contained even though the 
limitations imposed upon its length will force us to skip all the 
remarks on the history of the subject as well as on references and 
on a broader context. Fortunately, interested readers may very 
easily get acquainted with these aspects of the new theory in 
several very thorough and extensive reviews \cite{Carla} and also in 
our own recent compact review \cite{SIGMA} and/or in our two years 
old short contribution \cite{acta} to Acta Polytechnica.

In  section \ref{Csec} we shall start our discussion from the 
bound-state models characterized by the loss of observability of 
complexified ``coordinates". In the generic dynamical scenario where 
the Riemann surface of the wave functions can be assumed 
multisheeted we shall define certain monodromy-sensitive models 
called quantum toboggans. Our selection of their sample applications 
will cover innovative models possessing several branch points in the 
complex $x-$plane and/or exhibiting supersymmetry.  

Section \ref{Dsec} will offer information about the specific 
cryptohermitian approach to bound-state models characterized by the 
manifest time-dependence of their operators of observables (cf. 
paragraph \ref{D1sec}) or by the presence of a fundamental length in 
the theory (cf. paragraph \ref{D2sec}).

The two possible mechanisms of a return to unitarity in the models 
of scattering by complex potentials will finally be described in 
Section~\ref{Bsec}. Via concrete examples we shall emphasize there 
the beneficial role of a ``smearing" of phenomenological potentials 
and the necessity of an appropriate redefinition of the effective 
mass in certain regimes.

Section \ref{Fsec} contains a few concluding remarks. For the sake 
of completeness, a few technical remarks concerning the role of the 
Dyson mapping in the abstract formulation of Quantum Theory as well 
as in some of its concrete applications will be added in the form of 
three Appendices.

\section{Quantum theories working with
quadruplets of alternative Hilbert spaces\label{Csec}}


Within cryptohermitian approach a new category of models of bound 
states appeared, a few years ago, under the name of quantum 
toboggans \cite{2005}. Their introduction  extended the class of 
integration paths of complexified ``coordinates" $x=q(s)$ in the 
standard Schr\"{o}dinger equations to certain topologically 
nontrivial complex trajectories. The Hamiltonians 
$H^{(T)}=p^2+V^{(T)}(x)$  containing analytic potentials 
$V^{(T)}(x)$ with singularities (the superscripts $^{(T)}$ stand 
here for ``tobogganic") were connected with the generalized complex 
asymptotic boundary conditions and specified as operating in a 
suitable Hilbert space ${\cal H}^{(T)}$ of wave functions in which 
the Hamiltonian itself is manifestly non-Hermitian.

Practical phenomenological use of any cryptohermitian quantum model 
requires, firstly, the sufficiently persuasive demonstration of the 
reality of its spectrum and, secondly, the availability of at least 
one metric operator $\Theta=\Theta(H)$ (cf. Appendices A - C for its 
definition). Usually, both of these conditions are nontrivial so 
that any form of the solvability of the model is particularly 
helpful. {\it Vice versa}, once the Hamiltonian $H$ proves solvable 
in Hilbert space ${\cal H}^{(T)}$, we may rely upon the availability 
of the closed solutions of the underlying Schr\"{o}dinger equations 
and on the related specific spectral representations of the 
necessary operators (cf. \cite{pseudo,identifi} for more details).

The topological nontrivality of the tobogganic paths of coordinates 
running over several Riemann sheets of wave functions happened to 
lead to severe complications in the  numerical attempts to compute 
the spectra. This difficulty becomes almost insurmountable when the 
wave functions describing quantum toboggans happen to possess two or 
more branch points (cf. \cite{tob2} for an illustrative example). 
For these reasons it is recommended to rectify the tobogganic 
integration paths via a suitable change of variables in a 
preparatory step \cite{Novotny}. Our tobogganic Schroedinger 
equations then acquire the generalized eigenvalue-problem form 
$H\psi=E W\psi$ of the so called Sturm-Schroedinger equations with 
the rectified Hamiltonian $H\neq H^\dagger$ and with a nontrivial 
weight operator $W\neq W^\dagger \neq I$. Both of these operators 
are defined in another, transformed, ``more friendly" Hilbert space 
${\cal H}^{(F)}$ of course~\cite{tobrev}.

\subsection{Supersymmetric quantum toboggans \label{CBsec}}

\begin{figure}[h]                     
\begin{center}                         
\epsfig{file=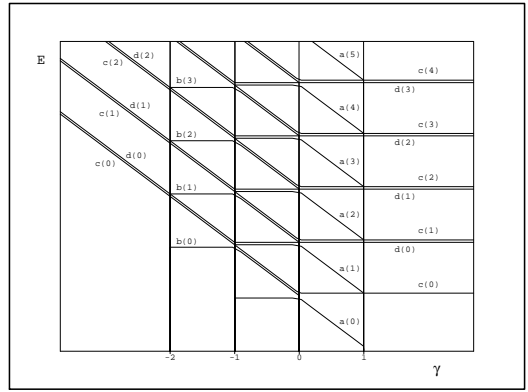,angle=270,width=0.5\textwidth}
\end{center}                         
\vspace{-2mm}\caption{Spectrum of singular supersymmetric harmonic 
oscillator.
 \label{fiedee}}
\end{figure}

The introduction of the  cryptohermitian and tobogganic models 
proved useful in the context of supersymmetry (SUSY). A large number 
of papers devoted to this subject exists. Their representative 
sample is referenced in \cite{Carla}. The easiest case (called 
supersymmetric quantum mechanics) uses just the Hamiltonian and  the 
two charge operators generating  SUSY algebra,
 \ben
{\cal H}= \left [
 \begin{array}{cc}
H^{(-)}&0\\ 0&H^{(+)}\ea \right ]= \left [
 \begin{array}{cc}
  B^{} A^{}
&0\\ 0& A^{} B^{} \ea \right ] \,,\ \ \ \ \ \  {\cal Q}=\left [
 \begin{array}{cc}
0&0\\
A^{}&0 \ea \right ], \ \ \ \ \ \ \tilde{\cal Q}=\left [
 \begin{array}{cc}
0& B^{}
\\
0&0 \ea \right ]\,.  \label{keyb}
 \een
For the solvable model of ref.~\cite{hosusy} the energy spectrum 
(composed of four families $E_n=a(n),\ldots, d(n)$) is displayed in 
figure \ref{fiedee}. At $\gamma=-1/2$ the singularity vanishes and 
the (up to the ground state) doubly degenerate SUSY spectrum becomes 
strictly equidistant.

The imposition of supersymmetry has been extended to quantum 
toboggans in \cite{SUSYtob}. Both the components of the 
super-Hamiltonian were defined along topologically nontrivial 
complex curves which connect several Riemann sheets of the wave 
function. The new feature of this generalized model lies in the 
non-uniqueness of the map ${\cal T}$ between "tobogganic" partner 
curves. As a consequence, we must redefine the creation- and 
annihilation-like operators as follows,
 \ben
  A=
 -{\cal T}\frac{d}{dx} + {\cal T}W^{(-)}(x)\,,\ \ \
 B=\frac{d}{dx}{\cal T}^{-1} + W^{(-)}(x){\cal T}^{-1}
 \,.
 \label{reredefinition}
 \een
In contrast to the non-tobogganic cases the Hermitian-conjugation 
operator ${\cal T}$ even ceases to be involutory (i.e., ${\cal 
T}\neq {\cal T}^{-1}$, cf. paper \cite{SUSYtob} for more details).

\subsection{Four-Hilbert-space Quantum Mechanics \label{C1sec}}

In a way explained in our papers~\cite{identifi} the tobogganic 
quantum systems with real energies generated by their apparently 
non-Hermitian Hamiltonians may be assigned the entirely standard and 
consistent probabilistic interpretation. This assignment is based on 
a replacement of the initial Hilbert space ${\cal H}^{(T)}$ by 
another, ``friendly" Hilbert space ${\cal H}^{(F)}$ in which the 
above-mentioned Sturm-Schroedinger equations $H\psi=E W\psi$  have 
to be solved. This forces us to replace the three-Hilbert-space 
scheme of paper \cite{SIGMA} [cf. also Appendices A and B and 
figure~\ref{fieeone}] by the following four-Hilbert-space pattern of 
mappings
 \ben
 \ba
 \begin{array}{|c|}
 \hline 
   {\bf  tobogganic}\   {\rm space\ }{\cal H}^{\rm (T)}\ \\
 \
  {\rm analytic\ {multivalued}}\ \psi[q(s)] \\ %
 \
  {\rm  multisheeted\ paths}\  q^{(N)}(s)\\
  \hline
 \ea\ \ \
 \ \ \ \ \ \ \ \ \ \ \ \ \ 
 \
 \begin{array}{|c|}
 \hline%
  {\rm {\bf  physics\ } in\ } {\cal H}^{\rm (P)}\
 \\  \ \mathfrak{h}=\mathfrak{h}^\dagger\,,
 \ \  \mathfrak{w}=\mathfrak{w}^\dagger\
    \\ %
  \ \
  {\rm dynamics\ via\ topology }\     \\ 
 \hline
 \ea
 \\
 \\
 \ \ \ \ \
  \downarrow
  \ \    \stackrel{\rm (the\ change\ of\ variables) }{\rm {\bf rectification\  }}
  \ \  \  \ \ \ \ \ \ \ \ \ \
  \ \  \  
  \ \  \  \ \ \ \
\ \ \ \ \   \uparrow\ \ \
 \stackrel{{\rm (the\ unitary\ mapping)} }{\rm \bf equivalence} \\
 \\
 \begin{array}{|c|}
 \hline%
  {\rm  {\bf  feasibility}\ in \
   } {\cal H}^{\rm (F)}\ \\
 \
  {\rm }\   H \neq H^\dagger\,,\ \ \ W \neq W^\dagger \\ %
   {\rm  Sturm-Schr\ddot{o}dinger\ eqs.}  \\
  \hline
 \ea
  \
  \stackrel{\longrightarrow }{
   \stackrel{\rm  (metric\ is\  introduced)}{{\bf hermitization }}}
 \
 \begin{array}{|c|}
 \hline%
  {\rm {\bf  standard\ }space\   } {\cal H}^{\rm (S)}\
 \\   \ H = H^\ddagger\,,\ \ \ W=W^\ddagger
    \\ ad\ hoc\ 
  {\rm metric  }\  \Theta\neq I \\ %
 \hline
 \ea
\\
\\
\ea
 \een
The analyticity of the original wave function $\psi[q(s)]$ along the 
given tobogganic integration path with parameter $s \in 
(-\infty,\infty)$ is assumed. The rectification transition between 
Hilbert spaces ${\cal H}^{(T)}$ and ${\cal H}^{(F)}$ is tractable as 
an equivalence transformation under this assumption \cite{tobrev}. 
In the subsequent sequence of maps $F \to S$ and $F \to P$ one 
simply follows the old three-Hilbert-space pattern of Appendix C 
\cite{SIGMA} in which just the nontrivial weight operators $W$ 
and/or $\mathfrak{w}$  are added and appear in the respective 
generalized Sturm-Schr\"{o}dinger equations. 

Marginally, let us add that the suitably modified spectral 
representations of the eligible metric operators may be used, say, 
in the form derived in \cite{identifi}. The purely kinematical and 
exactly solvable topologically nontrivial ``quantum knot" example of 
ref. \cite{knots} can also be recalled here as an exactly solvable 
illustration in which the confining role of the traditional 
potential is fully simulated by the mere topologically nontrivial 
shape of the complex integration path.


\section{Bound-state theories working with
the triplets of alternative Hilbert spaces\label{Dsec}}

\subsection{Quantum models admitting the time-dependence of their 
cryptohermitian  Hamiltonians \label{D1sec}}

In our review  \cite{SIGMA}  of the three-Hilbert-space (3HS) 
formalism we issued a warning that some of the consequences of the 
enhanced flexibility of the language and definitions may sound as 
new paradoxes. For illustration let us mention just that in the 3HS 
approach the generator $H_{\rm (gen)}=H_{\rm (gen)}(t)$ of the 
time-evolution of wave functions is allowed to be different from the 
Hamiltonian operator $H=H(t)$ of the system in 
question~\cite{timedep}.

The key to the disentanglement of the similar puzzles is easily 
found in the explicit specification of the Hilbert space in which we 
define the Hermitian conjugation. We showed in \cite{timedep} that 
the use of the full triplet of spaces of figure \ref{fieeone} 
becomes unavoidable  whenever our cryptohermitian observables are 
assumed time-dependent because their variations may and must be 
matched by the time-dependence of the representation of the physical 
{\em ad hoc} Hilbert space ${\cal H}^{(S)}$. Indeed, its nontrivial 
inner product is capable to play the role of a ``moving frame" image 
of the original physical Hilbert space ${\cal H}^{(P)}$. Although  
our ``true" Hamiltonian (i.e., operator $\mathfrak{h}(t)$ in  ${\cal 
H}^{(P)}$) is the generator of the time evolution in ${\cal 
H}^{(P)}$, the time-evolution of the wave functions  in ${\cal 
H}^{(S)}$ is controlled not only by the ``dynamical" influence of 
$H=H(t)$ itself but also by the ``kinematical" influence of the 
time-dependence of the ``rotating" Dyson mapping $\Omega=\Omega(t)$. 
Thus, the existence of any other {\em given and manifestly 
time-dependent} observable $\mathfrak{o}(t)$ in ${\cal H}^{(P)}$ 
will {\em leave its trace} in  Dyson map $\Omega(t)$, i.e., in 
metric $\Theta(t)$, i.e., in the time-dependence of the ``moving 
frame" Hilbert space ${\cal H}^{(S)}$. 

This circumstance implies the existence of two pullbacks of the 
evolution law from ${\cal H}^{(P)}$ to ${\cal H}^{(S)}$, with the 
recipe $|\varphi(t)\kt=\Omega^{-1}(t)\, |\varphi(t)\!\pkt$ being 
clearly different from the complementary recipe $\br\!\br 
\varphi(t)\,|=\pbr\!\varphi(t)\,|\,\Omega(t)$. The same Dyson 
mapping leads to the two different evolution operators, viz., to the 
evolution law for kets,
 \ben
 |\varphi(t)\kt=U_R(t)\, |\varphi(0)\kt\,,\ \ \ \ \ \
 U_R(t)=\Omega^{-1}(t)\,u(t)\,\Omega(0)
 \een
and to the different evolution law for brabras,
 \ben
 |\varphi(t)\kt\!\kt=U_L^\dagger(t)\, |\varphi(0)\kt\!\kt\,,\ \ \ \ \ \
 U_L^\dagger(t)=\Omega^\dagger(t)\,u(t)\,
 \left [\Omega^{-1}(0)\right ]^\dagger\,.
 \een
We have no space here for the detailed reproduction of the whole 
flow of this argument as presented in \cite{timedep}. Its final 
outcome is the definition of the {\em common} time-evolution 
generator 
 \[
 H_{(gen)}(t)=H(t) -{\rm i}\Omega^{-1}(t)
 \dot{\Omega}(t)\,.
 \]
entering the final doublet of time-dependent Schr\"{o}dinger 
equations
 \bea
 {\rm i}\partial_t|\Phi(t)\kt
 =H_{(gen)}(t)\,|\Phi(t)\kt\,,
 \label{SEA}\\
 {\rm i}\partial_t|\Phi(t)\kt\!\kt
 =H_{(gen)}(t)\,|\Phi(t)\kt\!\kt\,.
 \label{SEbe}
 \eea
This ultimately clarifies the artificial character and redundancy of 
the Mostafazadeh's conjecture \cite{plb} of quasistationarity, i.e., 
of the requirement of time-independence of the inner products and of 
the metric, i.e., {\em ipso facto}, of Hilbert space ${\cal 
H}^{(S)}$. 

\subsection{Systems admitting a controllable nonlocality 
\label{D2sec}}

In a way emphasized by Jones \cite{Jones} the direct observability 
of coordinates $x$ is lost for the majority of the 
parity-times-time-reversal-symmetric (or, briefly, ${\cal 
PT}-$symmetric) quantum Hamiltonians. In the context of scattering  
this forced us to admit a non-locality of the potentials in 
\cite{prd}. Fortunately, in the context of bound states the loss of 
the observability of coordinates is much less restrictive since we 
do not need to prepare any asymptotically free states. The 
admissible Hilbert-space metrics $\Theta$ may be then chosen as 
moderately non-local acquiring, in the simplest theoretical scenario 
as proposed in our paper \cite{fund}, the form of a short-ranged 
kernel in a double-integral normalization or in the inner products 
of the wave functions. Thus, the standard Dirac's delta-function 
kernel is simply reobtained in the zero-range limit.

In refs. \cite{prd,discrete} we proposed several bound-state toy 
models exhibiting, in a confined-motion dynamical regime, various 
forms of an explicit control of the measure $\theta$ of their 
dynamically generated non-locality. The exact solvability of some of 
these models even allowed us to assign each Hamiltonian the {\em 
complete} menu of its hermitizing metrics $\Theta=\Theta_\theta$ 
distinguished by their optional fundamental lengths $\theta\in 
(0,\infty)$. In this setting the local metrics reappear at 
$\theta=0$ while certain standard hermitizations only appeared there 
as infinitely long-ranged, with $\theta=\infty$.

\section{Scattering theories using pairs of Hilbert spaces 
${\cal H}^{(P)} \neq {\cal H}^{(F)}$ \label{Bsec}}

In our last illustrative application of 3HS formalism let us select 
just two non-equivalent Hilbert spaces  ${\cal H}^{(F,S)}$ and turn 
to scattering theory where one assumes that the coordinate is 
certainly measurable/measured at large distances. This means that we 
may employ the operators in coordinate representation and  accept 
only such models where the metric operator remains asymptotically 
proportional  to delta function,  $\br x|\Theta|x'\kt \sim 
\delta(x-x')$ at $|x|\gg 1$ and $|x'|\gg 1$. A few concrete models 
of this type were described in refs.~\cite{prd,discrete} using 
minimally nonlocal, ``smeared" point interactions of various types 
(which were, in the latter case, multi-centered). The use of 
nonperturbative discretization technique rendered possible the 
construction of the (incidentally, unique) metric $\Theta$ 
compatible with the required asymptotic locality.

The resulting physical picture of scattering was unitary and fully 
compatible with our intuitive expectations. In our last paper 
\cite{SIGMAlast} the scope of the theory has further been extended 
to the generalized scattering models where the matrix elements  $\br 
x|\Theta|x'\kt$ of the metric were allowed  operator-valued.

A slightly different approach to scattering has been initiated in 
paper \cite{Gezov} where we studied the analytic and ``realistic" 
Coulombic cryptohermitian potentials defined along  U-shaped complex 
trajectories circumventing the origin in the complex $x$ plane from 
below. Unfortunately, this model was unstable with respect to 
perturbations. A few years later we clarified, in paper 
\cite{negmass}, that a very convenient stabilization of the model 
may be based on a minus-sign choice of the bare mass in 
Schr\"{o}diner equation. Very soon afterwards we also revealed that 
the scattering by the amended Hamiltonian is unitary \cite{Coul}. 
The transmission and reflection coefficients were evaluated in  
closed analytic form exhibiting the coincidence of the bound-state 
energies with the poles of the transmission coefficients. Thus, 
after a moderate modification a number of observations forming the 
analytic theory of S-matrix has been found transferrable to the 
cryptohermitian quantum theory.

\section{Conclusions \label{Fsec}}

One of paradoxes characterizing Quantum Theory may be seen in a
contrast between its stable status in experiments (where, typically,
its first principles are appreciated as unexpectedly robust
\cite{Styer}) and its fragile status in mathematical context (where
virtually all of its rigorous formulations are steadily being found,
for this or that reason, not entirely satisfactory \cite{Strocchi}). 
In fact, at least a part of this apparent conflict is just a 
pseudoconflict. Its roots can be traced back to various purely 
conceptual misunderstandings. In our present review we emphasized 
that within the comparatively narrow framework of quantum  theory 
using cryptohermitian representations of observables the majority of 
these misunderstandings can be clarified, mostly via a careful use 
of an adequate notation.

The core of our present message can be seen in the unified outline 
of the resolution of the internet-mediated debate (cf. \cite{SIGMA} 
for references) in which the admissibility and  consistent 
tractability of the manifestly time-dependent cryptohermitian 
observables has been questioned. It is now clear that the reduction 
of the scope of the theory to the mere quasistationary systems as 
proposed by Mostafazadeh \cite{plb} is unfounded. 

This bound-state-related message can be seen accompanied by the 
clarification of a return to unitarity in the models of scattering 
mediated by cryptohermitian interactions. The currently valid 
conclusion is that it makes sense to combine the complexification of 
the short-range interactions with our making them at least slightly 
nonlocal. We saw that in parallel, also the metric can be required 
to exhibit a certain limited degree of nonlocality. 

New questions emerge in this context. This means that in spite of 
all the recent quick progress the current intensive development of 
the cryptohermitian quantum theory is still fairly far from its 
completion.

\section*{\label{Usec}Appendix A: Hilbert space in our present notation}

In our review paper  \cite{SIGMA} we explained that one of the most 
natural formulations of the abstract Quantum Theory should follow 
the ideas of Scholtz et al \cite{Geyer} by constructing the three 
parallel representatives of any given wave function living in the 
three separate Hilbert spaces. We argued that the use of the 
three-Hilbert-space (3HS) formulation of Quantum Theory seems best 
capable to clarify a few paradoxes emerging in connection with the 
concept of Hermiticity and encountered in the recent literature. We 
emphasized in \cite{SIGMA} that many quantum Hamiltonians with real 
spectra characterized by their authors as manifestly non-Hermitian 
{\em should and must} be re-classified as Hermitian. In this sense 
we fully accepted the dictum of standard textbooks on quantum theory 
and complemented the corresponding postulates just by a few 
explanatory comments.

In a brief summary of this argument let us recall that the states 
$\psi$ of a (say, one-dimensional) quantum system are often assumed 
represented by normalized elements of the simplest physical and 
computation friendly concrete Hilbert space 
$\mathbb{L}^2(\mathbb{R})$. This is already just a specific 
assumption with restrictive consequences. Thus, in a more ambitious 
picture of a general quantum system each state $\psi$ should only be 
perceived as an element $|\psi\kt$ of an {\em abstract} vector space 
${\cal V}$. The equally abstract {\em ``dual"} vector space ${\cal 
V}'$ of linear functionals over ${\cal V}$ may be then, in general, 
``bigger",  ${\cal V}'\supset {\cal V}$. In the most common selfdual 
case with  ${\cal V}'= {\cal V}$ one speaks about the  Hilbert space 
${\cal H}^{(F)}\ :=\ ({\cal V},{\cal V}')$ where the superscript 
$^{(F)}$ stands, say, for ``(user-)friendly" or ``feasible". 
 
In many standard formulations of the first principles of Quantum 
Theory the well known Dirac's bra-ket notation is being used, with 
$|\psi\,\kt \in {\cal V}$ and $ \br\,\psi| \in {\cal V}'$ for a {\em 
fixed} or ``favored" Hilbert space ${\cal H}^{(F)}$. At the same 
time, this choice of the notation does not exclude a transition 
(say, $\Omega$) to some other vector and Hilbert spaces denoting, 
e.g.,  $\Omega\,|\psi\kt :=|\psi\,\pkt \in {\cal W}$ using just the 
slightly deformed, ``spiked" kets~\cite{SIGMA}.

\section*{Appendix B: 
Dyson mapping $\Omega$ as a nonunitary  generalization of the 
Fourier transformation ${\cal F}$  \label{A2sec} }

In the context of nuclear physics the use of the single, favored 
Hilbert space ${\cal H}^{(F)}$ is rather restrictive. For example, 
in the context of the so called interacting boson model and in the 
way inspired by the well known advantages of the use of the usual 
unitary Fourier transformation ${\cal F}=\left [{\cal 
F}^\dagger\right ]^{-1}$, nuclear physicists discovered that their 
constructive purposes may be much better served by a suitable 
generalized, manifestly non-unitary (often called Dyson) invertible 
mapping $\Omega$. 

More details may be found in paper \cite{Geyer} where the operators 
$\Omega$ were described as mediating the transition from a friendly 
bosonic vector space ${\cal V}$ into another, fermionic and 
``physical" vector space ${\cal W}$. The deepened mathematical 
differences between ``bosonic" (i.e., simpler) ${\cal V}$  and  
fermionic (i.e., complicated, computationally much less accessible) 
${\cal W}$ weakens the parallelism between $\Omega$ and ${\cal F}$ 
since the latter operator merely switches between the so called 
coordinate- and momentum-representations of $\psi$s lying in {\em 
the same} Hilbert space $\mathbb{L}^2(\mathbb{R})$. 

This encouraged us to propose, in \cite{SIGMA}, the visual 
identification of the bras and kets in the one-to-one correspondence 
to the space in which they live, with $|\psi\,\kt \in {\cal V}$ 
while  $\Omega\,|\psi\kt :=|\psi\,\pkt \in {\cal W}$. For duals 
(i.e., bra-vectors) we recommended the same notation, with  $ 
\br\,\psi| \in {\cal V}'$ while $ \br \psi|\,\Omega^\dagger :=\pbr 
\psi| \in {\cal W}'$.

\section*{Appendix C: The connection between Dyson map $\Omega$ and 
metric $\Theta$\label{quattro}}

In the notation of Appendix B one represents the {\em same} state 
$\psi$ in {\em two non-equivalent} Hilbert spaces, viz., in the 
friendly F-space ${\cal H}^{(F)}\ :=\ ({\cal V},{\cal V}')$ and in 
the physical P-space ${\cal H}^{(P)}\ :=\ ({\cal W},{\cal W}')$  
(characterized by the ``spiked" kets and bras). The latter space is, 
by construction, manifestly non-equivalent to the former one  since, 
by definition, we have, for overlaps, $\pbr \psi_a|\psi_b\,\pkt=\br 
\psi_a|\,\Omega^\dagger \Omega\,|\psi_b\kt \neq \br 
\psi_a|\psi_b\kt$.

\begin{figure}[h]                     
\begin{center}                         
\epsfig{file=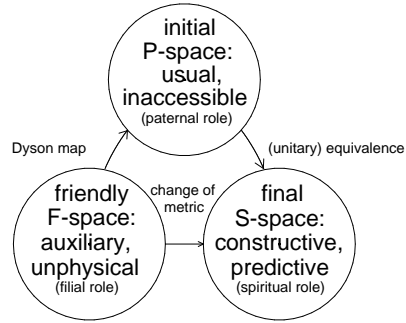,angle=270,width=0.5\textwidth}
\end{center}                         
\vspace{-2mm}\caption{The {\bf same}
 physics is predicted in ${\cal H}^{(P)}$ and in ${\cal H}^{(S)}$
 while, presumably, 
 calculations are all performed in ${\cal H}^{(F)}$.
 \label{fieeone}}
\end{figure}
According to our review \cite{SIGMA} the demonstration of unitary 
non-equivalence  between ${\cal H}^{(F)}$  and ${\cal H}^{(P)}$ can 
easily be converted into a proof of unitary equivalence between 
${\cal H}^{(P)}$ and {\em another}, third, ``standardized" Hilbert 
space  ${\cal H}^{(S)}\ :=\ ({\cal V},{\cal V}'')$. Indeed, we are 
free to introduce a redefined vector space of linear functionals 
${\cal V}''$  such that the equivalence will be achieved. For the 
latter purpose it is sufficient to introduce the special duals $ 
\bbr\,\psi| \in {\cal V}''$ denoted by the new, ``brabra" 
Dirac-inspired symbol. In terms of a given Dyson operator $\Omega$ 
we may define these brabras, for the sake of definiteness, by the 
formula $ \bbr\,\psi_a| =\br \psi_a|\,\Theta$ of ref.~\cite{pseudo} 
where we abbreviated $\Theta=\Omega^\dagger \Omega\,$. 

In \cite{Geyer} the new  operator $\Theta$ has been called metric. 
It defines the inner products in  the ``second auxiliary" (i.e., in 
its nuclear-physics exemplification, second bosonic) Hilbert space 
${\cal H}^{(S)}$ which is, by construction, unitarily equivalent to 
the original physical  ${\cal H}^{(P)}$. The whole 3HS scheme is 
given the compact graphical form in figure~\ref{fieeone}.

\subsection*{Acknowledgment}

This work has been supported by the M\v{S}MT Doppler Institute 
project LC06002, by the Institutional Research Plan AV0Z10480505 and 
by GA\v{C}R grant  202/07/1307.


\end{document}